\documentclass[
aps,%
10pt,%
final,%
notitlepage,%
oneside,%
twocolumn,%
nobibnotes,%
nofootinbib,%
superscriptaddress,%
showpacs,%
centertags,%
showkeys,%
amssymb]{revtex4}

\usepackage{graphicx}
\usepackage{amsmath}

\def\,{\ifmmode\mskip\thinmuskip\else\leavevmode\thinspace\fi}

\newcommand{\vecc}[1]{\mbox{\boldmath $#1$}}
\newcommand{\unit}[1]{\,\textmd{#1}}
\renewcommand{\Im}{\,\textmd{Im}}
\newcommand\m[1]{\mathrm {#1}}
\def\dd{\mathrm d}

\begin{document}

\title{Sum rule for a difference of proton and neutron total
photoproduction cross-sections}

\author{E.~Barto\v s} \affiliation{Joint Institute for
Nuclear Research, 141980 Dubna, Russia}

\author{S.~Dubni\v{c}ka } \affiliation{Inst.\ of Physics,
Slovak Acad.\ of\ Sci., D\'ubravsk\'a cesta 9, 845 11 Bratislava,
Slovakia}

\author{E.~A.~Kuraev} \affiliation{Joint Institute for Nuclear
Research, 141980 Dubna, Russia}


\begin{abstract}
Starting from very high energy inelastic electron-nucleon
scattering with a production of a hadronic state $X$ to be moved
closely to the direction of the initial nucleon, then utilizing
analytic properties of parts of forward virtual Compton scattering
amplitudes on proton and neutron, one obtains the relation between
nucleon form factors and a difference of proton and neutron
differential electroproduction cross-sections. In particular, for
the case of small transferred momenta, one finally derives sum
rule, relating Dirac proton mean square radius and anomalous
magnetic moments of proton and neutron to the integral over a
difference of the total proton and neutron photoproduction
cross-sections.
\end{abstract}

\maketitle

At the end of sixties of the last century Kurt Gottfried, by a
consideration of the very high-energy electron-proton scattering
and the nonrelativistic quark model of hadrons, has found
\cite{Gottfried}\footnote{We would like to thank S.~B.~Gerasimov
for bringing this paper to our attention.} a sum rule relating the
proton mean square charge radius $\langle r^2_{Ep}\rangle$ and the
proton magnetic moment $\mu_p = 1 + \kappa_p$ to the integral over
the total proton photoproduction cross-section
$\sigma_{tot}^{\gamma\m{p}}(\nu)$ in the form

\begin{equation} \label{eq:1}
\int\limits_{0}^{\infty} \frac{\dd \nu}{\nu}\sigma_{tot}^{\gamma
\m{p}}(\nu) = \frac{\pi^2\alpha}{m^2_p}\left[
\frac{4}{3}m^2_p\langle r^2_{Ep}\rangle +1-\mu^2_p\right],
\end{equation}
where $\nu$ is the energy loss in the laboratory frame, $\alpha$
is the fine structure constant and $m_p$ is the proton mass.

Nowadays it is well known, that the Gottfried sum rule cannot be
satisfied since the corresponding integral diverges due to the
known rise of the total proton photoproduction cross-section at
high energies.

In this paper by means of a distinct way from the Gottfried
approach and considering the nucleon isodoublet (proton and
neutron) simultaneously a new sum rule is derived, which relates
Dirac proton mean square radius and anomalous magnetic moments of
proton and neutron to the integral over a difference of the total
proton and neutron photoproduction cross-sections. Thus the rise
of both photoproduction cross-sections at high energies is
mutually cancelled and the corresponding integral converges.

In a derivation of the new sum rule we start also with a
consideration of the very high energy electron-nucleon scattering
\begin{equation} \label{eq:2}
e^-(p_1) + N(p) \rightarrow e^-(p_1') + X
\end{equation}
with a production of a hadronic state $X$ moving closely to the
direction of initial nucleon. The corresponding one photon
exchange approximation matrix element takes the form

\begin{equation}\label{eq:3}
M=i\frac{\sqrt{4\pi\alpha}}{q^2}\bar{u}(p_1')\gamma_\mu u(p _1)
\langle X\mid J_\nu^{QED}\mid N^{(r)}\rangle g^{\mu\nu}
\end{equation}
where $(r)$ means a spin-state of the nucleon.

If the Gribov representation \cite{Gribov} of the metric tensor in
the photon propagator of (\ref{eq:3}) is applied
\begin{equation}\label{eq:4}
g_{\mu\nu}=g_{\mu\nu}^{\bot}+\frac{2}{s}\big(\tilde{p}_{\mu}
\tilde{p}_{1\nu}+\tilde{p}_{\nu}\tilde{p}_{1\mu}\big)
 \approx\frac{2}{s} \tilde{p}_\mu\tilde{p}_{1\nu},\;
s=(p_1+p)^2
\end{equation}
where
\begin{equation}\label{eq:5}
\tilde{p}_1=p_1-m^2_ep/(2p_1p), \quad \tilde{p}=p-m^2_Np_1/(2p_1p)
\end{equation}
are almost light-like vectors to be used according to Sudakov
\cite{Sudakov} for an expansion of the photon transferred
fourmomentum $q$
\begin{eqnarray}\label{eq:6}
q=\beta_q \tilde{p}_1+\alpha_q\tilde{p}+q_\bot,\quad
q_\bot=(0,0,\vecc{q}),\\\nonumber
 \tilde{p}q_\bot=\tilde{p_1}q_\bot=0,
 \quad q_\bot^2=-\vecc{q}^2,\\
\nonumber
\end{eqnarray}
for the corresponding cross-section one obtains
\begin{multline}\label{eq:7}
\dd\sigma=\frac{4\pi\alpha}{s(q^2)^2}p_1^\mu
p_1^\nu\times\\
\sum_{X\neq N}\sum^{1/2}_{r=-1/2} \langle N^{(r)}\mid J_\mu^{QED}
\mid X\rangle^* \langle X\mid J_\nu^{QED} \mid N^{(r)} \rangle
\dd\Gamma,
\end{multline}
where summations through the created hadronic states $X$ and the
spin states of the initial nucleon are carried out and $\dd\Gamma$
denotes the final state phase-space volume. Further, approximating
square momentum of virtual photon
\begin{equation}\label{eq:8}
q^2\approx -[\vecc{q}^2+(m_es_1/s)^2]
\end{equation}
with
\begin{equation}\label{eq:9}
s_1= 2(qp)=s\beta_q,
\end{equation}
i.~e., it is related to the invariant square mass of the final
hadronic state by the relation
\begin{equation}\label{eq:10}
m_X^2 = s_1 + q^2 + m_N^2,
\end{equation}
and transforming the phase space volume of the final electron into
the form
\begin{equation}\label{eq:11}
\frac{1}{(2\pi)^3}\frac{\dd^3p'_1}{2\epsilon'_1}=\frac{1}{(2\pi)^3}\dd^4q
\delta[(p_1-q)^2]=\frac{1}{(2\pi)^3}\frac{\dd s_1}{2s}\dd^2q_\bot,
\end{equation}
one gets the final state phase-space volume in the form
\begin{align}\label{eq:12}
\dd\Gamma = \frac{\dd s_1}{2s(2\pi)^3}\dd^2{q_\bot}\dd\Gamma_X;
\end{align}
$$\dd\Gamma_X=(2\pi)^4 \delta^4(q+p-\sum_j p_j)\prod_j\frac{\dd^3p_j}
{2\varepsilon_j(2\pi)^3}.$$

Besides, using the current conservation condition
\begin{multline}\label{eq:13}
q^\mu \langle X\mid J_\mu^{QED}\mid N^{(r)}\rangle\\
\approx(\beta_qp_1+q_\bot)^\mu\langle X\mid J_\mu^{QED}\mid
N^{(r)}\rangle = 0,
\end{multline}
($\alpha_q\tilde p$ gives a negligible contribution) one can write

\begin{multline}\label{eq:14}
\int p_1^\mu p_1^\nu \sum_{X\neq N} \sum^{1/2}_{r=-1/2}\langle
N^{(r)}\mid J_\mu^{QED}\mid X\rangle^* \times \\
\langle X \mid J_\nu^{QED}\mid N^{(r)}\rangle  \dd \Gamma_X= p_1^\mu p_1^\nu
\Delta \tilde {A}^{(N)}_{\mu \nu }=\\
\frac{s^2}{s_1^2}\vecc{q}^2e^ie^j \Delta \tilde {A}^{(N)}_{ij}=
2i\frac{s^2}{s_1^2}\vecc {q^2} \Im \tilde {A}^{(N)}(s_1,\vecc{q}),
\quad e_i=\frac{\vecc{q}_i}{|\vecc{q}|},
\end{multline}
where just Cutkosky rules for $s$-channel discontinuity $\Delta
\tilde{A}^{(N)} =2i \Im \tilde {A}^{(N)}$ of the corresponding
Feynman amplitude was applied.

Here we would like to note that the amplitude
$\tilde{A}(s_1,\vecc{q})$ by a construction is only a part of the
total forward virtual Compton scattering amplitude
$A(s_1,\vecc{q})$, which doesn't contain (unlike the amplitude
$A(s_1,\vecc{q})$) any crossing Feynman diagram contributions. As
a result there is no $u$-channel pole in
$\tilde{A}(s_1,\vecc{q})$, which is a crucial point in a
derivation of the new sum rule by using the analytic properties of
the latter amplitude.

Since the imaginary part of the crossing Feynman diagrams is
starting to be different from zero only above
\begin{equation}\label{eq:15}
s_1^{(3N)}=8m^2_N+\vecc{q}^2,
\end{equation}
one can write down an equality relation
\begin{equation}\label{eq:16}
\Im \tilde{A}(s_1,\vecc{q})= \Im A(s_1,\vecc{q})= 4
s_1\sigma_{tot}^{\gamma^* \m{p}\to X}(s_1,\vecc{q}),
\end{equation}
for $2m_Nm_\pi+m_\pi^2+\vecc{q}^2\leq s_1\leq 8m_N^2+\vecc{q}^2$.
Fortunately above the threshold (\ref{eq:15}) total proton and
neutron photoproduction cross-sections are almost equal and at the
new sum rule one can integrate over them up to infinity.

Now, substituting relation (\ref{eq:14}) into cross-section
expression (\ref{eq:7}) and integrating over the phase space
volume of the final hadronic state $X$, as well as over the
invariant mass squared $m^2_X$, i.~e., over the variable $s_1$
(see(\ref{eq:10})), to be interested only for $\vecc{q}$
distribution, for a difference of corresponding differential
proton and neutron electroproduction cross-sections one finds (see
Appendix D in \cite{Baier})

\begin{multline} \label{eq:17}
\left(\displaystyle\frac{\dd\sigma^{\m{e}^-\m{p}\to
\m{e}^-X}(s,\vecc{q})}
{\dd^2\vecc{q}}-\displaystyle\frac{\dd\sigma^{\m{e}^-\m{n}\to
\m{e}^-X}(s,\vecc{q})} {\dd^2\vecc{q}}\right )=\\
\frac{\alpha\vecc{q}^2}{4\pi^2}
\displaystyle\int\limits_{2m_Nm_\pi+m_\pi^2+\vecc{q}^2}^{\infty}
\frac{\dd s_1}{s_1^2[\vecc{q}^2+(m_es_1/s)^2]^2}\times\\
\left (\Im \tilde{A}^{(p)}(s_1,\vecc{q})-\Im
\tilde{A}^{(n)}(s_1,\vecc{q})\right ).
\end{multline}

If one neglects the second term in square brackets of the
denominator of the integral of (\ref{eq:17}) (due to the small
value of $m_e$ and high $s$ in comparison with $s_1$) and takes
into account (\ref{eq:16}) for $\vecc{q}^2\to 0$ together with the
relation $\dd^2\vecc{q}=\pi \dd\vecc{q}^2$ one comes to the
expression
\begin{multline} \label{eq:18}
\vecc{q}^2\left(\frac{\dd\sigma^{\m{e}^- \m{p}\to \m{e}^-
X}}{\dd\vecc{q}^2}-\frac{\dd\sigma^{\m{e}^- \m{n}\to \m{e}^-
X}}{\dd\vecc{q}^2}\right)\Big |_{\vecc{q}^2\to 0} =\\
\frac{\alpha}{\pi}\int\limits_{2m_Nm_\pi+m_\pi^2}^\infty \frac{\dd
s_1}{s_1}\left(\sigma_{tot}^{\gamma \m{p}\to
X}(s_1)-\sigma_{tot}^{\gamma \m{n}\to X}(s_1)\right) ,
\end{multline}
similar to the difference of the total cross-sections of the
electroproduction processes on proton and neutron in the
Weizs\"{a}cker--Williams approximation \cite{W1,W2}
\begin{multline}\label{eq:19}
\left(\sigma_{tot}^{\m{e}^-\m{p}\to \m{e}^-X}(s) -
\sigma_{tot}^{\m{e}^-\m{n}\to \m{e}^-X}(s)\right) =\\
2\frac{\alpha}{\pi}\ln\left(\frac{s}{m_em_\pi}\right)\times\\
\int\limits_{2m_Nm_\pi+m_\pi^2}^\infty \frac{\dd
s_1}{s_1}\left(\sigma_{tot}^{\gamma \m{p}\to X}(s_1)-
\sigma_{tot}^{\gamma \m{n}\to X}(s_1)\right) .
\end{multline}

\begin{figure*}[hbt]
\includegraphics[scale=.6]{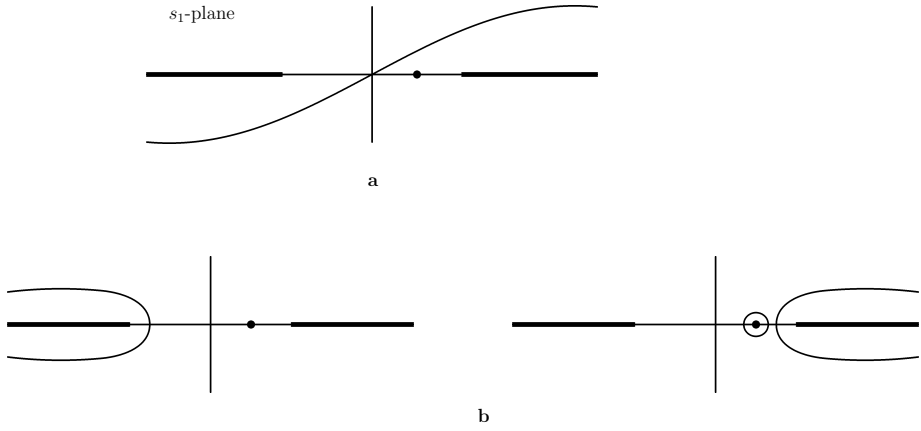}
\caption {\label{fig:1}Sum rule interpretation in $s_1$ plane. On
subfigure (a) is drawn the contour C [see (20)], on the subfigure
(b) is the contour C closed to upper and lower half plane.}
\end{figure*}

The analytic properties of the amplitude $\tilde{A}(s_1,\vecc{q})$
in $s_1$-plane are consisting of the one-nucleon intermediate
state pole at $s_1=\vecc{q}^2$, the right-hand cut starting at the
pion-nucleon threshold $s_1=2m_Nm_\pi+m_\pi^2+\vecc{q}^2$ and the
left-hand cut starting from $s_1=-\vecc{q}^2-8m_N^2$.

If one defines the path integral $I$ in $s_1$-plane as presented
in Fig.~\ref{fig:1}a
\begin{equation}\label{eq:20}
I=\int\limits_C \dd s_1 \frac{p_1^\mu p_1^\nu}{s^2}
\left(\tilde{A}^{(p)}_{\mu\nu}(s_1,\vecc{q})-
\tilde{A}^{(n)}_{\mu\nu}(s_1,\vecc{q})\right )
\end{equation}
then once the contour $C$ is closed to upper half-plane, another
one to lower half-plane (Fig.~\ref{fig:1}b) and considering
(\ref{eq:14}), the following sum rule

\begin{multline}\label{eq:21}
\pi\left(\textmd{Res}^{(n)}-\textmd{Res}^{(p)}\right)=\\
\vecc{q}^2\int\limits_{r.h.}^\infty\frac{\dd s_1} {s_1^2}
\left(\Im \tilde{A}^{(p)}(s_1,\vecc{q})- \Im
\tilde{A}^{(n)}(s_1,\vecc{q})\right)
\end{multline}
appears with (an averaging through the initial nucleon and photon
spins is performed)
\begin{equation}\label{eq:22}
 Res^{(N)}={2\pi\alpha}\left(F^2_{1N}+ \frac
{\vecc{q}^2}{4 m^2_N}{F^2_{2N}}\right)
\end{equation}
to be the one-nucleon intermediate state pole contribution and the
left-hand ($l.h.$) cut contributions from the difference $\left(
\Im \tilde{A}^{(p)}-\Im \tilde{A}^{(n)}\right)$ are mutually
annulated.

Substituting (\ref{eq:22}) into (\ref{eq:21}) and taking into
account (\ref{eq:17}) one comes to the relation
\begin{multline} \label{eq:23}
F_{1n}^2(-\vecc{q}^2)+\frac{\vecc{q}^2}{4m_n^2}
F_{2n}^2(-\vecc{q}^2)\\
 -F_{1p}^2(-\vecc{q}^2)-\frac{\vecc{q}^2}{4m_p^2} F_{2p}^2(-\vecc{q}^2)=\\
2\frac{(\vecc{q}^2)^2}{\pi\alpha^2} \left(\frac{\dd\sigma^{\m{e}^-
\m{p} \to \m{e}^-X}}{\dd\vecc{q}^2}-\frac{\dd\sigma^{\m{e}^- \m{n}
\to \m{e}^-X}}{\dd\vecc{q}^2}\right ).
\end{multline}
For $\vecc{q}^2=0$ the right-hand side is equal to zero, but the
left-hand side is $-1$. This is caused by the following reasons.
On the right-hand side we take into account only strong
interactions effects and on the left-hand side the $-1$ is given
by nonzero proton charge, i.~e., pure electromagnetic effect. In
order to separate the pure strong interactions from
electromagnetic ones on the left-hand side of the sum rule one has
to regularize the latter by adding $+1$ in order to achieve the
zero also there. As a result the sum rule takes the final form
\begin{multline}\label{eq:24}
1 + F_{1n}^2(-\vecc{q}^2)+\frac{\vecc{q}^2}{4m_n^2}
F_{2n}^2(-\vecc{q}^2)\\
-F_{1p}^2(-\vecc{q}^2)-\frac{\vecc{q}^2}{4m_p^2} F_{2p}^2(-\vecc{q}^2)=\\
2\frac{(\vecc{q}^2)^2}{\pi\alpha^2} \left(\frac{\dd\sigma^{\m{e}^-
\m{p} \to \m{e}^-X}}{\dd\vecc{q}^2}-\frac{\dd\sigma^{\m{e}^- \m{n}
\to \m{e}^-X}}{\dd\vecc{q}^2}\right ),
\end{multline}
giving into a relation nucleon electromagnetic form factors with a
difference of the differential cross-sections of deep inelastic
electron-proton scattering. There is a challenge to specialized
experimental groups (e.~g., in DESY) to verify the sum rule
(\ref{eq:24}).

Now, taking a derivative of both sides in (\ref{eq:24}) according
to $\vecc{q}^2$ for $\vecc{q}^2\to 0$ and employing the relation
(\ref{eq:18}) one comes to the new sum rule relating Dirac proton
mean square radius and anomalous magnetic moments of proton and
neutron
\begin{equation}\label{eq:25}
\langle r_{1p}^2\rangle =
6\frac{\dd}{\dd{q}^2}F_{1p}(q^2)\Big|_{q^2=0},\quad
\kappa_{N}=F_{2N}({q}^2)\Big|_{q^2=0}
\end{equation}
to the integral over a difference of the total proton and neutron
photoproduction cross-sections (we used for laboratory frame
$s_1=2M_N\omega$)
\begin{multline}\label{eq:26}
\frac{1}{3}\langle r_{1p}^2\rangle
-\frac{\kappa_p^2}{4m_p^2}+\frac{\kappa_n^2}{4m_n^2}=\\
\frac{2}{\pi^2\alpha}\int\limits_{\omega_N}^\infty
\frac{\dd\omega} {\omega}\left[\sigma_{tot}^{\gamma \m{p}\to
X}(\omega)- \sigma_{tot}^{\gamma \m{n}\to X}(\omega)\right]
\end{multline}
with $\omega_N=m_\pi+m_\pi^2/2M_N$, in which just a mutual
cancellation of the rise of these total proton and neutron
photoproduction cross-sections for $\omega\to \infty$, created by
the Pomeron exchanges, is achieved. It is straightforward to see
that this final sum rule is not influenced at all by a
renormalization of the left-hand side in (\ref{eq:24}) by $+1$.

Using the Dirac proton mean square radius from \cite{Dubnicka1}
and proton and neutron anomalous magnetic moments from
\cite{Hagiwara}, the evaluation of the left-hand side in
(\ref{eq:26}) gives $(1.93\pm 0.18)\unit{mbarn}$.

On the other hand the data on photoproduction cross section on the
neutron are not so known as for the proton case up to now.
Nevertheless, taking compilation of both of them from
\cite{Baldini}, and assuming that both total cross-sections are
starting at the pion mass and are equal above the last neutron
experimental point at $\omega = 17.84 \unit{GeV}$ one gets on the
right-hand side the value $(1.92 \pm 0.32)\unit{mbarn}$. So, the
sum rule (\ref{eq:26}) can be considered to be satisfied.

\begin{acknowledgments}
We are grateful to Lev Lipatov for valuable discussions and
participants of the seminar at BLTP JINR, Dubna for comments. One
of the authors (E.A.K) is grateful to ECT (Trento) Centrum. E.A.K
and E.B would like to thank to Institute of Physics SAS in
Bratislava for a warm hospitality.

The work was in part supported by Slovak Grant Agency for
Sciences, Gr.\ No.\ 2/1111/23, E.A.K. is grateful to INTAS grant
00366.
\end{acknowledgments}

\end{document}